\documentclass [12pt] {article}
\begin{document}
\newcommand{\be}{\begin{equation}}
\newcommand{\ee}{\end{equation}}
\vspace{2.0 cm}
\begin{center}
\begin{Large}
A generalization of Mach's principle:\\
effect of the velocity relative to surrounding matter\\
\end{Large}
\vspace{0.3 in}
Sergio L. Dagach\\
BECH, Alameda 1111, Santiago, Chile\\
\vspace{0.1 in}
Luis A. Dagach\\
Miguel Claro 239, Providencia, Santiago, Chile\\
\end{center}
\vspace{0.1 in}

\noindent
We postulate that \begin{itshape}all the presently known kinematic effects on physical quantities related to a material particle (e.g., masss increase, time dilation, variation of the field of a charge) are due to its velocity relative to surrounding matter, and not to the observer's reference frame. The minimal velocity (i.e., the velocity that minimizes these quantities) relative to a single large body being a function of the distance to and mass of the body.\end{itshape} In consequence, the minimal velocity is a function of position, and the reference frame associated to this velocity is strictly of local validity. In this local frame, quantities related to a particle are independent of its direction of motion. We further assume that, at any given point, light propagates isotropically solely in the minimal-velocity local frame existing at the point. We obtain the following results: (i) After showing the compatibility of the gravitational field eqs. with our assumptions, we find the functional dependance of the minimal velocity on the distance to and mass of a single large body. (ii) A permanent gravitational field is the convective rate of change of the minimal velocity field. (iii) A Lorentz transformation connects the values of quantities related to a particle, for two different velocities of the particle relative to its minimal-velocity local frame. However, a Lorentz transformation does not connect this frame with any other moving uniformly with respect to it. (iv) \begin{itshape}The experimentally detected effects of kinematic, as well as gravitational, mass increase and time dilation are derived. This is, they all are due to the presence of the nearby (single) large mass.\end{itshape} (v) Fizeau's experiment, Michelson's experiment, aberration of fixed stars are taken account of.  (vi) Contrary to what special relativity predicts, a Michelson's experiment performed from an inertial vehicle orbiting the earth, or the sun, should detect the orbital velocity of the vehicle.


\thispagestyle{empty}

\addtocounter{page}{-1}
\newpage
\section{Introduction}

   The theory of special relativity, in dealing with uniform motion, establishes the physical equivalence of all Galilean frames. With this theory, fundamental physical quantities related to a material particle become dependent on the velocity of the particle relative to the observer's inertial system. The theory of general relativity establishes that the dynamic ef\mbox{}fects arising from acceleration are actually due to the acceleration relative to the surrounding matter as a whole.\footnote{ We are not making here a point on to what extent Mach's principle is contained in Einstein's theory of general relativity.} This latter theory introduces then a dif\mbox{}ferent type of relativity, showing that fundamental physical quantities related to a material particle are, in addition, dependent on the gravitational field present, i.e., on the distribution of the surrounding matter.

   We believe that the current dual character of motion and physical quantities related to matter is unnecessary, and that, to a great extent, it originated from the fact that by the time Mach introduced his hypothesis (that the inertial ef\mbox{}fects arising from rotation are actually due to rotation relative to the surrounding matter), the (classical) principle of relativity already represented a fundamental axiom of the scientif\mbox{}ic thought. In this way, any possible ef\mbox{}fect on physical quantities related to a material particle due to its velocity relative to the surrounding matter was never pondered. Furthermore, such a hypothesis would have lacked experimental support, since no physical velocity-dependent phenomenon that could be ascribed to the velocity of material particles relative to the surrounding matter was known at that time. However, if inertial ef\mbox{}fects arise in a material particle being subject to an acceleration (i.e.,  to a modif\mbox{}ication of its velocity) relative to the surrounding matter, it is reasonable to assume that the value itself of the inertia of the particle (as well as the values of other physical quantities related to the particle) might actually be dependent on its velocity relative to the surrounding matter.

   If the velocity of a material particle relative to the surrounding matter happened, somehow, to determine the values of physical quantities related to the particle, we should be able to interpret as ef\mbox{}fects caused by this relative velocity all those phenomena presently interpreted as phenomena caused by the velocity of the particle relative to the observer's frame of reference, and, in consequence, the principle of relativity would become superfluous.\footnote{However, the concept of a Newtonian absolute space would still remain invalid.} It is an experimental fact that observers at rest on the earth routinely detect a mass increase of highly accelerated material particles. However, it is also a fact that these observers happen to have close by the mass of the earth at all times, such a way that the velocity, relative to these observers, of a material particle moving in the vicinity of the earth coincides with the velocity of the particle relative to the earth itself. Thus, observers at rest on the earth might be actually measuring a mass increase due to this latter velocity, that is, a mass increase independent of the observer's frame of reference. We will postulate: \begin{itshape}physical quantities related to a material particle are dependent on its velocity relative to surrounding matter. The minimal velocity relative to a single large body is a function of the distance to and mass of the body.\end{itshape}.

The minimal velocity (``Machian velocity"), at any given point, is the velocity relative to the surrounding matter of a material particle located at the point, such that it minimizes (or maximizes) physical quantities related to the particle. According to the above postulate, the Machian velocity is a function of position and (because of the relative motion between the large masses of the universe) time. If the surrounding matter consists of just a single celestial body, the Machian velocity at any given point will be a certain velocity relative to this body, which is a function of the distance to and mass of the body. In the presence of multiple celestial bodies, the Machian velocity at any given point should be the result of the combined effect of all these bodies.\footnote{We are implicitly assuming that the Machian velocity is independent of the material particle we use and the physical quantity we measure.}

A specific local frame is associated to the Machian velocity existing at any given point (``Machian local frame''). This is the local frame that has the minimal velocity relative to the surrounding matter, i.e., a frame which is strictly valid only at the point (and instant) considered. Only in the Machian local frame that exists at a particle's momentary position, physical quantities related to the particle are independent of its direction of motion. We will refer as "local velocity" to the velocity of a particle relative to the Machian local frame that exists at its momentary position.\footnote{Thus, it is not that the whole matter of the universe provides with a unique frame of reference, valid everywhere, respect to which motion would have a physical meaning.} 

   According to the previous analysis, a \begin{itshape}velocity f\mbox{}ield\end{itshape} exists throughout all of space, which at every point is determined by the distribution of the surrounding matter. We will refer as ``Machian space'' to the deformable frame of reference whose velocity, at any given point, corresponds to the Machian velocity.  Being the Machian velocity a function of position, we will have at any given point a gradient of this velocity, and consequently, an acceleration f\mbox{}ield, which, in general, is also function of position. \begin{itshape}We identify this acceleration f\mbox{}ield with the gravitational f\mbox{}ield caused by matter\end{itshape}: as it moves throughout a region of space (where we have introduced a system of coordinates), a force-free material particle keeps a constant velocity relative to the Machian local frame existing at its momentary position. Being the velocity of this local frame a function of position, the particle accelerates relative to our system of coordinates. Thus, in this scheme, ``real forces'' means those forces that accelerate a material particle relative to the Machian local frame existing at the particle's momentary position, and ``f\mbox{}ictitious forces'' those forces that arise from the acceleration of the observer's frame of reference relative to that Machian local frame. Gravitational forces correspond then to f\mbox{}ictitious forces.

   Our postulate, specific to the electromagnetic phenomena, becomes: \begin{itshape}An electromagnetic field is dependent on the local velocity of the sources.\end{itshape} It is then reasonable to assume: \begin{itshape}At any given point in space, light propagates isotropically solely in the Machian local frame existing at the given point.\end{itshape} Thus, light does not propagate isotropically in any other local frame, including those that move uniformly with respect to the Machian local frame (these are local inertial frames, in the sense that in them law of inertia is valid). In consequence, systems of coordinates placed in different local inertial frames are not connected by a Lorentz transformation. However, in section 5, we will show that the values of physical quantities related to a material particle, for two different  local velocities of the particle, are connected by a Lorentz transformation.

   In section 2 we determine the Machian velocity in terms of the distance to and mass of a single celestial body. In section 3 we derive the functional dependence of physical quantities related to a material particle on the local velocity of the particle. In section 4 we show that the experimentally detected velocity and gravity dependancies of physical quantities related to a material particle turn out both to be manifestations of the dependance of these quantities on the local velocity of the particle. In section 5 we introduce coordinative definit\mbox{}ions compatible with our hypotheses, and we derive the corresponding laws of transformation connecting systems of coordinates placed in different local inertial frames. In section 6 we take account of some classical ef\mbox{}fects, which were signif\mbox{}icant in the turn of the century, namely, Fizeau's experiment, Michelson's experiment, aberration of fixed stars. We also show in this section that the propagation of light throughout a deformable space requires a hypothesis concerning its path. In section 7 we propose an experiment to test this scheme.

\section{Determination of the Machian velocity}

   We will incorporate into this scheme some of the concepts and the mathematical formalism employed by the general theory of relativity, with the following observations:
\begin{enumerate}
\item Our above conclusion that gravitational forces are f\mbox{}ictitious forces, corresponding to the principle of equivalence, emerges here as a natural consequence of our starting postulate.

\item Several authors \cite{Fock:Weinberg} have stressed the real meaning of the principle of general covariance and its relation to the principle of equivalence. It is known that practically any theory can be expressed in a generally covariant manner (i.e., independent of the particular system of coordinates in which they are expressed), and that the covariance by itself cannot lead to any physical consequences. The real meaning of this principle is to identify the gravitational f\mbox{}ield with the geometrical properties of space-time, what is in line with our previous analysis.

\item The general theory of relativity introduces the principle of local validity of special relativity, this is, it assumes the physical equivalence of the local inertial frames.\footnote{Actually, this principle is customarily introduced by def\mbox{}ining the local inertial frames as those local frames where the metric tensor assumes the constant form, in local pseudo-Cartesian coordinates introduced in the frames, these local coordinates being then connected by Lorentz transformations.} However, it is also known that this principle is not essential in a generally covariant theory of gravitation. In the words of Weinberg \cite{Fock:Weinberg}: \begin{quote}In particular, general covariance does not imply Lorentz invariance -there are generally covariant theories of gravitation that allow the construction of inertial frames at any point in a gravitational f\mbox{}ield, but that satisfy Galilean relativity rather than special relativity in these frames.\end{quote} In our scheme, at any given point, a preferred local inertial frame is physically identif\mbox{}iable. Hence, essentially, our hypotheses replace the principle of local validity of special relativity.

\item Einstein's f\mbox{}ield equations are independent of the particular hypothesis that one introduces about the physical character of the local inertial frames. They def\mbox{}ine the dependance of the curvature of space-time on the surrounding matter. In our scheme, the curvature of space-time amounts to the state of motion of the Machian space, basically, its velocity from point to point. We adopt Einstein's f\mbox{}ield equations to describe this state of motion.
\end{enumerate}

   In what follows, we will use the convention that Latin indices run from 1 to 4 and Greek indices run from 1 to 3. Let us introduce an arbitrary system S of coordinates $(x^{i})$ in four-space, with $x^{i} = (x,y,z,ct)$, valid in a f\mbox{}inite region of space. Space-time is a manifold with a Riemannian structure, i.e.,

\be
ds^{2} = g_{ik}dx^{i}dx^{k} \; .
 \ee

   It is proved that in the immediate vicinity of any given point in a Riemannian manifold there always can be introduced a system of coordinates in which the metric tensor takes the normal form (locally geodesic coordinates) \cite{Tulio}. In space-time, the locally geodesic coordinates correspond to local pseudo-Cartesian coordinates S$^{0}\ (X,Y,Z,cT)$ introduced into a certain local frame of reference I$^{0}$.  The law of inertia is valid in I$^{0}$ as well as in any other local frame of reference that moves uniformly with respect to I$^{0}$. In this sense, all these local frames of reference are inertial frames. However, the form that the metric tensor takes in pseudo-Cartesian coordinates introduced in these frames is a matter of physical hypothesis. According to our initial postulate, I$^{0}$ is unique and it corresponds to the Machian local frame at the given point.\footnote{At this point, we deviate from the general relativity, which assumes the local validity of special relativity.}

	Let $w_{\iota}(x^{\kappa},t)$ be the velocity of the Machian local frame at the point P$(x^{\kappa})$, at the time $t$. The space and time dependance of this velocity implies the existence of an acceleration of the Machian space, which, in general, is also a function of space and time. This acceleration is given by

\be
a_{\iota} = \frac{\partial w_{\iota}} {\partial t} + w_{\kappa} \frac {\partial w_{\iota}} {\partial x^{\kappa}} \; .
\ee

\noindent
On the other side, a force-free material particle placed at the point P acquires an acceleration given by [3]

\be
a_{\iota} = - \frac{\partial \chi} {\partial x^{\iota}} - c^{*} \frac{\partial \gamma_{\iota}} {\partial t}\;,
\ee

\noindent
with

\be
c^{*} = c\sqrt{1 + \frac{2\chi} {c^{2}}} \; ,
\ee

\noindent
where $\chi$ and $\gamma_{\iota}$ are the gravitational scalar potential and vector potential, respectively. The local velocity of the force-free material particle remains constant as it moves from one point where the Machian space has a certain Machian velocity to another point where the Machian space has a dif\mbox{}ferent Machian velocity. Thus, Eq.\ (3) gives the acceleration of the Machian space at the point P. Matching Eqs.\ (2) and (3), and considering only the stationary case,

\be
a_{\iota} = w_{\kappa} \frac {\partial w_{\iota}} {\partial x^{\kappa}} = - \frac{\partial \chi} {\partial x^{\iota}}  \; .
\ee

\noindent
If only one component $w_{\iota}$ is non-null, say, $w_{1}$, one gets

\be
w_{1} = \sqrt{- 2\chi} \; .
\ee

In the case of a gravitational f\mbox{}ield created by a single concentrated mass $M$ located at the origin of our system of coordinates, and according to Schwarzschild's solution of Einstein's f\mbox{}ield equations,

\be
R_{ij} - \frac{1} {2}g_{ij}R = - \frac{8\pi G}{c^{4}}T_{ij} \; ,
\ee

\noindent
we have $\chi = -GM/r$ (with $G$ the gravitational constant). Introducing this value into Eq.\ (6), one gets that the velocity of the Machian space at a distance $r$ from the mass $M$ is given by 

\begin{equation} \label{eq:Machian velocity}
w = w_{r} = \sqrt{\frac{2GM} {r}} \; ,
\end{equation}

\noindent
which points towards the concentrated mass.\footnote{This result coincides, as expected, with the escape velocity.}

The acceleration of the Machian space at the distance $r$ is given by

\be
a = a_{r} = \vec{w} \cdot \nabla \vec{w} = - \frac{GM}{r^{2}} \; .
\ee

\noindent 
Thus, the acceleration of the Machian space, i.e., the gravitational f\mbox{}ield, around a large mass essentially arises from the gradient of velocity of the Machian space itself.

\section {Determination of the dependance of physical quantities related to a material particle on the local velocity of the particle}
 
   The time tracks of a freely falling material particle and of a light ray is a geodesic in space-time, i.e.,

\be
\delta \int \! ds = 0 \; .
\ee

\noindent
The equations of the geodesic obtained from this principle are given by

\be
\frac{d^{2}x^{i}} {ds^{2}} + \Gamma^{i}_{kj} \frac{dx^{k}}{ds} \frac{dx^{j}}{ds} = 0 \; ,
\ee

\noindent
with $\Gamma^{i}_{kj}$  the Christof\mbox{}fel symbols.  An integral for Eq.\  (11) is

\be
ds^{2} = constant \; .
\ee

\noindent
From Eq.\  (12), one gets

\be
dt = \Gamma d\tau \; ,
\ee

\noindent
with $\tau$ the proper time of the particle (time given by a standard clock that follows the particle in its motion), and

\be
\Gamma = \left\{\left[\left(1+\frac{2\chi}{c^{2}}\right)^{\frac{1}{2}}-\frac{\gamma_{\iota }u^{\iota}}{c} \right]^{2}-\frac{u^{2}}{c^{2}}\right\}^{-\frac{1}{2}} \; ,
\ee

\be
\chi = - \frac{c^{2}}{2} \left(1 + g_{44} \right) \; ,
\ee

\be
u^{\iota} = \frac{dx^{\iota}}{dt} \:, \;\; u = \left( \gamma_{\iota\kappa}u^{\iota}u^{\kappa} \right)^{\frac{1}{2}}  \; ,
\ee

\be
\gamma_{\iota} = \frac{g_{\iota4}}{\sqrt{\left(-g_{44}\right)}} \:, \;\; \gamma_{\iota\kappa} = g_{\iota\kappa} + \gamma_{\iota}\gamma_{\kappa} \;,
\ee

\noindent
where $u^{\iota}(x^{\iota},x^{4})$ and $u$ are the velocity and the speed of the particle relative to the arbitrary system S of coordinates, respectively, and $\gamma_{\iota\kappa}$ is the spatial metric tensor \cite{Moller}. Equation (13) gives the rate of a coordinate clock at rest in S compared with the rate of a moving standard clock.

   We introduce local pseudo-Cartesian coordinates S$^{0}\ (X,Y,Z,cT)$ into the Machian local frame I$^{0}$, at the point where the material particle is momentarily positioned. Referred to these coordinates, Eq.\ (12) becomes

\be
ds^{2} = dX^{2} + dY^{2} + dZ^{2} - c^{2}dT^{2} = constant \; . 
\ee

When the particle is at rest in I$^{0}$, Eq.\ (18) becomes

\be
ds^{2} = - c^{2}dT_{0}^{2} \; . 
\ee

\noindent
And the time T$_{0}$, given by coordinate clocks at rest in I$^{0}$, coincides with the proper time of the particle.

Referred to S$^{0}$, Eq.\ (13) becomes

\be
dT = \gamma \cdot dT_{0} \;,
\ee

\noindent
with

\be
\gamma \equiv \frac {1} {\sqrt{1 - \frac{U^{2}}{c^{2}}}} \; ,
\ee

\noindent
where

\be
U = \frac{\sqrt{dX^{2}+dY^{2}+dZ^{2}}} {dT}
\ee

\noindent
is the local speed of the particle (speed relative to I$^{0}$). Eq.\ (20) shows the dependance of the rate of a clock on its local speed. \begin{itshape}This time dilation is an ef\mbox{}fect independent of the velocity of the clock relative to the observer's frame of reference\end{itshape}.

\vspace {0.15 in }
   Let us assume we have a light clock, that is, a measuring-rod and a light ray reflecting repeatedly between its two end points A and B, with the measuring-rod laying in the direction of its local velocity $U$. The time that a ray of light takes in a round trip between the points A and B is given by

\be
dT = \frac{2dL/c}{\left( 1 - \frac{U^{2}}{c^{2}} \right)} \;,
\ee

\noindent
with $dL$ the length of the measuring-rod. Because of its local speed, the light clock will slow down, in the form given by Eq.\ (20). Comparing Eqs.\ (20) and (23), we get

\be
dL = \frac{dL_{0}}{\gamma} \;,
\ee

\noindent
with $dL_{0}$ the length of the measuring-rod when $U = 0$. Eq.\ (24) shows that the measuring-rod suf\mbox{}fers a contraction due to its local speed. As with the time dilation, this contraction is an ef\mbox{}fect independent of the velocity of the measuring-rod relative to the observer's frame of reference.

\vspace {0.15 in}
 Fundamental physical quantities related to a material particle are dependent on the local speed of the particle. In particular, this is the case with the duration of any physical process linked to the particle. Hence, the Machian space and the flow of time determined by the motion of the particle throughout this space are the space and time that pertains to the particle. A combination of them def\mbox{}ines the 4-momentum of the particle.

   Referred to the local pseudo-Cartesian coordinates S$^{0}$ introduced into the Machian local frame I$^{0}$ at its momentary location, a material particle travels a distance $dX^{i}$ in the direction $X^{i}$ in a time $d\tau$, as measured by a clock that moves along with the particle. The 4-momentum of the particle is def\mbox{}ined as

\be
 P_{i} = m_{0} \cdot \frac{dX^{i}}{d\tau} \;,
\ee

\noindent
with $m_{0}$ a constant, characteristic of the particle. From Eq.\ (25), we obtain

\be
 P_{\kappa} = m \cdot U_{\kappa} \;, \;\; P_{4} = m \cdot c = \frac{E}{c} \;,
\ee

\noindent
with

\be
m = \frac {m_{0}} {\sqrt{1 - \frac{U^{2}}{c^{2}}}} \;,
\ee

\noindent
where $m$ and $E$ are the mass and the total energy of the particle, respectively, and $U_{\kappa}$ its local velocity. Eq.\ (27) shows the dependance of the mass of a material particle on its local speed. Again, this mass increase is an ef\mbox{}fect independent of the velocity of the particle relative to the observer's frame of reference.

\section{Interpretation of the experimentally detected velocity and gravity dependancies of physical quantities related to a material particle}

	Schwarzschild's solution of the gravitational f\mbox{}ield equations applies in the vicinity of the earth, in an earth centered inertial frame, S$^{(e)}$, (i.e., in a system of coordinates which is attached to the c.\ of m.\ of the earth but does not follow the earth in its rotation relative to the distant stars). In this frame of reference, the velocity of the Machian space at a distance $r$ from the origin is given by Eq.\ (\ref{eq:Machian velocity}), with $M$ the mass of the earth. At the surface of the earth, we have $w^{2}/c^{2} \approx 10^{-9}$. Therefore, when a material particle moves in the vicinity of the earth with a local speed $U$ such that $U^{2}/c^{2} \gg 10^{-9}$, for all practical purposes, we can consider the Machian space at rest with respect to the earth at every point in this region. In this case, $U$ coincides with the velocity of the particle relative to observers at rest on the earth, $v_{r}$. For a clock that moves along with the particle, we will have

\be
dT = \frac {dT_{0}} {\sqrt{1 - \frac{U^{2}}{c^{2}}}} \cong \frac {dT_{0}} {\sqrt{1 - \frac{v_{r}^{2}}{c^{2}}}} \; .
\ee

\noindent
And the mass of the particle will be given by

\be
m = \frac {m_{0}} {\sqrt{1 - \frac{U^{2}}{c^{2}}}} \cong \frac {m_{0}} {\sqrt{1 - \frac{v_{r}^{2}}{c^{2}}}} \; .
\ee

\noindent
The ef\mbox{}fects shown by Eqs.\ (28) and (29) are routinely detected and interpreted as relativistic ef\mbox{}fects, that is, as ef\mbox{}fects due to the relative velocity $v_{r}$.

   A clock that is placed at rest in S$^{(e)}$ at a height $r$ has a local speed corresponding to the speed of the Machian space at that height.  Because of this speed, the clock slows down, in the form given by Eq.\ (20), i.e.,

\be
dT = \frac {dT_{0}} {\sqrt{1 - \frac{U^{2}}{c^{2}}}} = \frac {dT_{0}} {\sqrt{1 - \frac{(2GM/c^{2})}{r}}} \; .
\ee

\noindent
Eq.\ (30) corresponds to the so-called gravitational time dilation.

   In summary, the experimentally detected velocity and gravity dependancies of physical quantities related to a material particle turn out to be manifestations of a same effect, namely, dependance of these quantities on the local velocity of the particle.

\section{Coordinative def\mbox{}initions. Laws of transformation}

   Let W$^{(P)}$ be a region around a point P, of dimensions such that, for all practical purposes, we can consider the Machian local frame at P valid in W$^{(P)}$ (the size of W$^{(P)}$ being determined by the distances and masses of the nearby celestial bodies). Let I$^{0}$ be the Machian local frame at the point P, and I$^{1}$ an inertial local frame, introduced at the same point P, which moves with an arbitrary constant velocity $v_{1}$ relative to I$^{0}$. Let S$^{0}$ and S$^{1}$ be pseudo-Cartesian coordinates in I$^{0}$ and I$^{1}$, respectively. The transformation of coordinates that connect S$^{0}$ and S$^{1}$ interpret the hypotheses that we have formulated about the concepts of space and time. We have concluded that the rate of a clock and the length of a measuring-rod placed in W$^{(P)}$ are dependent on their velocities relative to I$^{0}$. Furthermore, in W$^{(P)}$, light propagates with the same velocity in all directions solely in I$^{0}$.

   We place standard measuring-rods and clocks at dif\mbox{}ferent points in W$^{(P)}$, at rest in I$^{0}$. The clocks, as well as the measuring-rods, are of identical construction, that is, the clocks show the same rate when placed next to each other, and the end-points of the measuring-rods coincide when placed next to each other. We assume that these instruments remain identical when placed at rest but at dif\mbox{}ferent points in W$^{(P)}$. We make use of the fact that light propagates isotropically in I$^{0}$ to synchronize these clocks: A light signal is emitted from a point A when the standard clock at this point records the time $t$. When this signal arrives at a point B, the standard clock at this point is set to $t + (dl/c)$, with $dl$ the distance from A to B and $c$ the speed of light. This synchronization is independent of the points A and B in W$^{(P)}$ and the time $t$ selected.

   We now place standard clocks and measuring-rods at dif\mbox{}ferent points in W$^{(P)}$, at rest in I$^{1}$, which are identical to the standard clocks and measuring-rods used in I$^{0}$. For the sake of simplicity, we assume that when the clock at rest in I$^{0}$ at the origin of S$^{0}$ shows zero, the origins of S$^{0}$ and S$^{1}$ coincide. We further assume that the axes in S$^{0}$ and S$^{1}$ are parallel to each other and that the velocity $v_{1}$ points along the X-axis. The clocks in I$^{1}$ are synchronized by putting them to zero when clocks at rest in I$^{0}$ and next to them show zero.\footnote{As it has been pointed out by dif\mbox{}ferent authors, and these def\mbox{}initions show, the def\mbox{}initional character of simultaneity does not exclude the eventual existence of a preferred inertial frame, here one of local character} We def\mbox{}ine the length of a moving line-segment, i.e., its moving-length, as ``the distance between simultaneous projections of its end points''. We assume that the moving-length of a line-segment is independent of the state of motion of the observer.

   Because of their local speed, clocks in S$^{1}$ slow down and measuring-rods contract. Thus, if $(X,Y,Z,T)$ and $(x_{1},y_{1},z_{1},t_{1})$ are the coordinates of a point event in S$^{0}$ and S$^{1}$, respectively, these coordinates will be related by

\be
x_{1} = \gamma_{1} \left[ X - v_{1}T \right] \:, \;\; y_{1} =Y \:, \;\;  z_{1} =Z \:, \;\; t_{1} =  \frac{T}{\gamma_{1}} \:, 
\ee

\noindent
and the inverse relations

\be
X = \frac{x_{1}}{\gamma_{1}} + \gamma_{1}v_{1}t_{1} \:, \;\; Y = y_{1} \:, \;\; Z = z_{1} \:, \;\; T = \gamma_{1}t_{1} \:, 
\ee

\noindent
where $\gamma_{1}$ corresponds to the expression given by Eq.\ (21), with $U$ replaced by $v_{1}$.

   We introduce now a frame of reference I$^{2}$, similar to I$^{1}$ but moving with a velocity $v_{2}$ relative to I$^{0}$, along the X-axis. If S$^{2}$ are pseudo-Cartesian coordinates in I$^{2}$, the transformations connecting S$^{1}$ and S$^{2}$ are given by

\be
x_{1} = \left( \frac{\gamma_{1}}{\gamma_{2}} \right)x_{2} + \gamma_{1}\gamma_{2} \left( v_{1} - v_{2} \right)t_{2} \:,\; y_{1} = y_{2}  \:,\; z_{1} = z_{2}  \:,\; t_{1} =  \left( \frac{\gamma_{2}}{\gamma_{1}} \right)t_{2} \: .
\ee

If a material particle is moving in the x-direction, with velocity $U$ in S$^{0}$, $u_{1}$ in S$^{1}$, and $u_{2}$ in S$^{2}$, the velocity transformation between S$^{0}$ and S$^{1}$, and between S$^{0}$ and S$^{2}$, are, respectively,

\be
u_{1} = \gamma_{1}^{2} \left(U \pm v_{1} \right) \:,\;\; u_{2} = \gamma_{2}^{2} \left(U \pm v_{2} \right) \: .
\ee

\noindent
And the velocity transformation between S$^{1}$ and S$^{2}$ is given by

\be
u_{1}=\gamma_{1}^{2} \left[\frac{u_{2}}{\gamma_{2}^{2}}-\left( v_{1} - v_{2} \right) \right] \; .
\ee 

   A set of physical quantities related to a material particle satisfy certain system equations. We know the values of these quantities when the local velocity of the particle is null. We can determine the values of these quantities when the particle has a non-null local velocity through the following procedure:

\begin{enumerate}
\item The particle is located at the point P and moves with a local velocity $v_{1}$, which is identical to the velocity of the frame of reference I$^{1}$ defined above.

\item Temporarily, we impose that the physical laws take in S$^{1}$ the same form as in S$^{0}$ and that light propagates also isotropically in I$^{1}$.

\item We look for transformations of coordinates and physical quantities between S$^{0}$ and S$^{1}$ such that they satisfy the requirements imposed in 2. Naturally, these transformations correspond to the Lorentz transformations.

\item We identify I$^{1}$ with I$^{0}$. We obtain in this way the values of the physical quantities when the system is in motion in I$^{0}$.

\end{enumerate}

   We see that in this context, the values of physical quantities related to a material particle, for two different local velocities of the particle, are connected by the Lorentz transformations. The Lorentz transformations for the coordinates do not connect two local inertial frames. At the present, observers at rest on the earth (i.e., approximately at rest in I$^{0}$), in order to determine the values of physical quantities of a moving system, follow the above procedure, giving to the Lorentz transformations for the coordinates the relativistic interpretation.

\section{Analysis of classical ef\mbox{}fects}

   In the nineteenth century, the existence of a medium which would be the carrier of the electromagnetic radiation, the ether, was pondered. The ether would permeate all the space and penetrate material media. As concerning with its state of motion, Michelson's experiment showed that everything happened as if the ether remained at rest with respect to the earth centered inertial frame S$^{(e)}$ (at least close to the surface of the earth)\footnote{ Michelson's experiment did not detect the orbital motion, but the rotation of the earth.}; Fizeau's experiment showed that the ether was partly dragged within the refringent media; and the aberration phenomenon discovered by Bradley (and later, Airy's experiment) showed that everything happened as if it was not dragged at all.

   For distances close enough to its surface, the earth can be considered as a single celestial body. Thus, everywhere in this region, and referred to S$^{(e)}$, the only motion of the Machian space is the vertical velocity given by Eq.\ (\ref{eq:Machian velocity}), that is, a motion that is independent of any motion of translation of the earth around the sun or the distant stars. This conclusion of\mbox{}fers an immediate explanation for the Michelson's experiment.

   As concerning with Fizeau's experiment, Lorentz's electron theory showed that the phase velocity of the light within a moving refractive medium is given, to a f\mbox{}irst approximation, by Fresnel's formula. To derive this result, Maxwell's equations (in their extended form to consider the presence of moving matter) are referred to a frame of reference that is at rest with respect to the observer, being assumed that this frame is not dragged at all by the refractive medium. This is the case of the Machian space in the vicinity of the earth: In this particular frame, Maxwell's equations take their simplest form, and its state of motion is determined fundamentally by the earth, independent of the motion of large distant celestial bodies or close matter of virtually negligible mass (compared to the mass of the earth).

   Because of the translation motion of the earth around the sun, there will be a gradual detachment of the Machian space around the earth, and an increasing attachment to a sun centered system of coordinates. Thus, the Machian space presents an additional velocity relative to the earth, opposite to the velocity of translation of the earth. The value of this additional velocity increases in the vertical direction, from essentially zero at the surface of the earth, up to the speed of translation of the earth (at a certain height $R$ which should be small compared to the distance to the sun), $V_{e}$. We wish to analyze the ef\mbox{}fect of the horizontal component of this velocity on the trajectory of a light ray which propagates throughout such a space, until it reaches the surface of the earth. This analysis amounts to analyze the behavior of the trajectory of a light ray when it crosses between two contiguous Machian local frames of inf\mbox{}initesimal thickness, which slide one with respect to the other at a relative speed $dv$, and then integrating these ef\mbox{}fects from $R$ to zero.

   In Figures 1 and 2, the Machian local frame I is at rest on the page and the Machian local frame I' moves with a velocity $dv$ to the right. A light ray propagates in I along the AO direction, as seen by observers in I (``proper trajectory'' in I), and, because of the velocity $dv$, along the BO direction as seen by observers in I'. We wish to determine the proper trajectory in I' (the trajectory in I' as seen by the observers in I is found by combining the velocity $c$, along the proper trajectory in I', with the velocity $dv$). In this respect, there is not any physical requirement that enforces a def\mbox{}inite answer for the proper trajectory in I'. In principle, one has two reasonable choices:
\begin{enumerate}
\item The proper trajectory in I' is the prolongation of the proper trajectory in I, i.e., the direction OA' in Fig.\ 1. The trajectories that observers in I' and in I see in I', are then given by OA' and OC', respectively. In summary, observers in I', as well as in I (or, for the same token, in any other frame of reference) see that the trajectory of the light ray suf\mbox{}fers a break at the crossing point O. 

\item The proper trajectory in I' is the prolongation of the trajectory that the observers in I' see in I, i.e., the direction OB' in Fig.\ 2. The trajectory that observers in I' and in I see in I' are then given by OB' and OA', respectively (the later trajectory corresponds, keeping only terms of f\mbox{}irst order in $dv/c$, to the prolongation of the trajectory AO). In summary, observers in I, as well as those in I', see that the trajectory of the light ray do not suf\mbox{}fer any break at the crossing point O. A similar ef\mbox{}fect is seen from any other frame of reference.
\end{enumerate}

   \begin{itshape}We will assume alternative 2 as the valid one.\end{itshape} In consequence, if the proper trajectory of a light ray at the height  $R$ makes an angle $\alpha$ with the direction of motion of the earth, then observers at rest on the surface of the earth will see the light ray making an angle  $\beta$ with this direction, where

\be
\tan \beta = \frac{\sin \alpha}{\cos \alpha + \frac{V_{e}}{c}} \;.
\ee

\noindent
As the light ray travels downward throughout Machian local frames of inf\mbox{}initesimal thickness, these observers will keep observing the light ray in the same direction, until it reaches them.\footnote{Regardless of the dependance of the horizontal component of the velocity of the Machian space  with the height.} The angle $\beta$ given by Eq.\ (36) corresponds to the angle of aberration.\footnote{One might try to establish an analogy with acoustical waves which follow alternative 1. However, it has to be kept in mind that these waves are of a different nature, and, in a f\mbox{}inal analysis, they result from the motion of the molecules of a material medium relative to a Machian local frame, which is the same in the two sections of the medium that would be sliding one with respect to the other.}
  
\section{Conclusions. Proposed experiment}

 In this paper we have focused our ef\mbox{}forts to establish the foundations of a scheme based on a matter-to-matter velocity dependance. We have postulated that physical quantities related to a material particle are dependent on its velocity relative to surrounding matter. The minimal velocity  (i.e., the velocity that minimizes these quantities) relative to a single large body is a function of the distance to and mass of the body. In consequence, the frame of reference associated to the minimal velocity, at any given point, has strictly a local validity. In the minimal-velocity local frame existing at a particle's momentary position, physical quantities related to the particle are independent of its direction of motion. Although our postulate discards an intrinsic dependance of physical quantities related to a material particle on its velocity relative to the observer's frame of reference, an observer would have to take into account the ef\mbox{}fect that his own velocity relative to the surrounding matter has on his measuring instruments. We have further assumed that, at any given point, light propagates isotropically solely in the minimal-velocity local frame existing at the point. We have found the dependance of the minimal velocity as a function of the distance to and mass of a single celestial body. We showed that a permanent gravitational field, at any given point, is the convective rate of change of the minimal velocity field. The experimentally detected velocity and gravity dependancies of physical quantities related to a material particle are both actually manifestations of the single dependance we have postulated. The values of physical quantities related to a material particle, for two different velocities of the particle relative to the minimal-velocity local frame existing at its momentary position, are connected by a Lorentz transformation. However, systems of coordinates placed in different local inertial frames are not connected by a Lorentz transformation.

   At the present, Lorentz invariance guides the construction of practically every physical theory. A revision of these theories would have to be done, proved that the scheme hereby proposed is experimentally validated. The fact that an electromagnetic f\mbox{}ield is dependent on the local velocity of its sources suggests that this field could also corresponds to some kind of motion of the Machian space around the sources (possibly, along with other fundamental new hypotheses). The distribution of this motion would be dependent on the velocity of the sources relative to the surrounding matter. It could be illuminating to perform Michelson's experiments in the presence of electromagnetic f\mbox{}ields with moving sources.
   
   The analysis made in section 6 for the trajectory of light rays propagating throughout a deformable space could be extended to the trajectory of material particles. The application of our alternative 2 to the trajectory of a force-free material particle that crosses between two contiguous Machian local frames, which are sliding one with respect to the other, results in a rectilinear motion of the particle.

   In the vicinity of the earth, the Machian space has essentially a vertical velocity in an earth centered inertial frame, i.e., it follows the motion of translation of the earth. Therefore, if we would perform a Michelson's experiment from a vehicle orbiting the earth, we should detect the orbital velocity of the vehicle. A similar statement is applicable for a vehicle orbiting the sun, far enough from any large mass.

\end{document}